\documentclass[twocolumn,showpacs,preprintnumbers,amsmath,amssymb,aps]{revtex4}

\usepackage{graphicx}% Include figure files
\usepackage{dcolumn}% Align table columns on decimal point
\usepackage{bm}% bold math

\begin{document}

\title{How To Attain Maximum Profit In Minority Game?}

\author{H. F. Chau and F. K. Chow}
\affiliation{
  Department of Physics, University of Hong Kong, Pokfulam Road, Hong Kong
}
\date{\today}

\begin{abstract}
What is the physical origin of player cooperation in minority game? And
how to obtain maximum global wealth in minority game? We answer the above
questions by studying a variant of minority game from which players choose
among $N_c$ alternatives according to strategies picked from a restricted set
of strategy space. Our numerical experiment concludes that player cooperation
is the result of a suitable size of sampling in the available strategy
space. Hence, the overall performance of the game can be improved by suitably
adjusting the strategy space size. 
\end{abstract}


\maketitle

Econophysics --- the study of economic and economic inspired problems by
physical means --- is the result of interflow between theoretical economists
and physicists. Using statistical mechanical and nonlinear physical methods,
econophysicists study global behaviors of simple-minded models of economic
systems making up of adaptive agents with inductive reasoning.
In particular, minority game (MG) \cite{Min1,Min2} is an important and
perhaps the most extensively studied econophysics model of global collective 
behavior in a free market economy. This game was proposed by Challet and 
Zhang under the inspiration of the El Farol bar problem introduced by the 
theoretical economist Arthur \cite{Min3}.

MG is a toy model of $N$ inductive reasoning players who have to choose one
out of two alternatives independently according to their best working
strategies in each turn. Those who end up in the minority side (that is, the
choice with the least number of players) win. Although its rules are
remarkably simple, MG shows a surprisingly rich self-organized collective
behavior. For example, there is a second phase transition between a symmetric
and an asymmetric phase \cite{Min4,Min5,Min6}.
Since the dynamics of MG minimizes a global function related to market
predictability, we may regard MG as a disordered spin glass system
\cite{Min7,Min8}.
Recently, Hart \emph{et al.} introduced the so-called crowd-anticrowd theory
to explain the dynamics of MG \cite{Min9,Min10}. Their theory stated that
fluctuations arised in the MG is controlled by the interplay between crowds
of like-minded agents and their perfectly anti-correlated partners.
The crowd-anticrowd theory not only can explain global behavior of MG, it
also provides a simple working hypothesis to understand the mechanism of
a number of models extended from the MG.

Numerical simulation as well as the crowd-antiwcrowd theory tell us that the
global behavior of MG depends on two factors. The first one is the product
of the number of players $N$ at play and the number of strategies $S$ each
player has. The second factor is the complexity of each strategy measured
by $2^{M+1}$, where $M$ is the number of the most recent historical outcomes 
that a strategy depends on. Global cooperation, as indicated by the fact that
average number of players winning the game each time is larger than the case
when all players make their choice randomly, is observed whenever $2^{M+1}
\approx N S$ \cite{Min4,Min5,Min6}. In fact, cooperative phenomenon is also
seen in our recent generalization of the MG in which each player can choose
one out of $N_c$ alternatives. More precisely, $N_c^M \approx N S$ is a
necessary condition for global cooperation between players in our
generalization \cite{Min11}.

Perhaps the two most important questions to address are why and when the
players cooperate in MG. In fact, these are the questions that the
crowd-anticrowd theory was trying to answer.
On the way of finding out the answers, Cavagna believed that the only
non-trivial relevant parameter to the dynamics of MG is $M$ \cite{Min12}.
But later on, Challet and Marsili revealed that historical outcomes also
determine the dynamics of MG in general. They also found that information
contained in the historical outcomes is irrelevant in the symmetric phase
\cite{Min13}.

Is it true that global behavior of MG is determined once $N$, $S$ and $M$
are fixed? More specifically, we ask if it is possible to lock the system
in a global cooperative phase for any fixed values of $N$, $S$ and $M$. In
this way, players, on average, gain most out of the game.
In what follows, we report a simple and elegant way to alter the complexity
of each strategy in MG with fixed $N$, $S$ and $M$. By doing so, it is
possible to keep (almost) optimal cooperation amongst the players in almost
the entire parameter space.

We begin our analysis by first constructing a model of MG with $N_c$
alternatives whose strategy space size equals $N_c^2$ for a fixed
prime power $N_c$. We label, for simplicity, the $N_c$ alternatives as the
$N_c$ distinct elements in the finite field $GF(N_c)$; and we denote this
variation of MG by MG($N_c$,$N_c^2$).
In MG($N_c$,$N_c^2$), each of the $N$ players is assigned once and for all
$S$ randomly chosen strategies. Each player then chooses one out of the
$N_c$ alternatives independently according to his/her best working strategy
in each turn. The choice chosen by the least \emph{non-zero} number of
players is the minority choice of that turn. (In case of a tie, the minority
choice is chosen randomly amongst the choices with least non-zero number of
players.)
The minority choice of each turn is announced. The wealth of those players
who end up in the minority side is added one point while the wealth of all
other players is subtracted by one.

To evaluate the performance of each strategy, a player uses the virtual
score which is the hypothetical profit for using that strategy in playing
the game. The strategy with the highest virtual score is considered as the
best performing one. (In case of a tie, one chooses randomly amongst those
strategies with highest virtual score.)
The only public information available to the players is the output of the
last $M$ steps.
A strategy $s$ can be represented by a vector $\vec{s} \equiv
(s_1, s_2, s_3, \ldots , s_L)$ where $L\equiv N_c^M$ and $s_i$ are the
choices of the strategy $s$ corresponding to different combination of the
output of the last $M$ steps.
In MG($N_c$,$N_c^2$), strategies are picked from the strategy space
${\mathbb S} = \{ \lambda_a \vec{v}_a + \lambda_u \vec{v}_u : \lambda_a,
\lambda_u \in GF(N_c) \}$ of size $N_c^2$ where $GF(N_c)$ denotes the finite
field of $N_c$ elements and all arithmetical operations are performed in the
field $GF(N_c)$. The two spanning strategy vectors $\vec{v}_a \equiv
( v_{a1}, v_{a2}, \ldots , v_{aL} )$ and $\vec{v}_u \equiv ( v_{u1}, v_{u2},
\ldots , v_{uL} )$ of the linear space ${\mathbb S}$ satisfy the following
two technical conditions:
\begin{equation}
 v_{ai} \neq 0 \mbox{~for~all~} i, \label{E:Cond_anti}
\end{equation}
and by regarding $i$ as a uniform random variable between $1$ and $L$,
\begin{equation}
 \mbox{Pr} (v_{ui} = k | v_{ai} = j) = 1/N_c \mbox{~for~all~} j,k \in GF(N_c)
 \label{E:Cond_uncorr}
\end{equation}
whenever $\mbox{Pr} (v_{ai} = j) \neq 0$. (We remark that these two technical
conditions are satisfied by various choices of $\vec{v}_a$ and $\vec{v}_u$
such as $v_{ai} = 1$ and $v_{ui} = f(i \bmod {N_c})$ where $f$ is a bijection
from ${\mathbb Z}_{N_c}$ to $GF(N_c)$.)

The span of the strategy vector $\vec{v}_a$ over $GF(N_c)$ forms a mutually
anti-correlated strategy ensemble ${\mathbb S}_a$ since
Eq.~(\ref{E:Cond_anti}) implies that any two distinct strategies drawn from
${\mathbb S}_a$ always choose different alternatives for any given historical
outcomes. Hence, the Hamming distance between any distinct strategies
$\vec{u}_1 \neq \vec{u}_2$ in ${\mathbb S}_a$ equals
\begin{equation}
 d ( \vec{u}_1, \vec{u}_2 ) = L. \label{E:Hamming_anti}
\end{equation}

In contrast, the span of the strategy vector $\vec{v}_u$ over $GF(N_c)$ forms
a mutually uncorrelated strategy ensemble ${\mathbb S}_u$ since 
Eq.~(\ref{E:Cond_uncorr}) and the fact that $\lambda\,GF(N_c) = GF(N_c)$ for
all $\lambda \in GF(N_c) \backslash \{ 0 \}$ imply that any two distinct
strategies drawn from ${\mathbb S}_u$ always choose their alternatives
independently for any given historical outcomes. In other words, the
probability that any two distinct strategies drawn from ${\mathbb S}_u$
choose the same alternative is equal to $1/N_c$. Consequently,
\begin{equation}
 d ( \vec{u}_3, \vec{u}_4 ) = L (1 - 1/N_c) \label{E:Hamming_uncorr}
\end{equation}
for any $\vec{u}_3 \neq \vec{u}_4 \in {\mathbb S}_u$; and
\begin{equation}
 d ( \vec{u}_1, \vec{u}_3 ) = L (1 - 1/N_c) \label{E:Hamming_uncorr2}
\end{equation}
for any $\vec{u}_1 \in {\mathbb S}_a$ and $\vec{u}_3 \in {\mathbb S}_u
\backslash \{ (0,0,\ldots ,0) \}$.

More generally, using Eqs.~(\ref{E:Hamming_anti})--(\ref{E:Hamming_uncorr2})
as well as the fact that $d(a,b) = d(a+c,b+c)$, we have
\begin{eqnarray}
& & d( \lambda_{a1} \vec{v}_a + \lambda_{u1} \vec{v}_u, \lambda_{a2} 
\vec{v}_a + \lambda_{u2} \vec{v}_u ) \nonumber \\
& = & d( [\lambda_{a1} - \lambda_{a2} ] \,\vec{v}_a, [\lambda_{u2} -
 \lambda_{u1} ] \,\vec{v}_u ) \nonumber \\
& = & \left\{ \begin{array}{ll}
 L (1-1/N_c) & \mbox{if~} \lambda_{u1} \neq \lambda_{u2}, \\
 L & \mbox{if~} \lambda_{u1} = \lambda_{u2} \mbox{~and~} \lambda_{a1} \neq
  \lambda_{a2}, \\
 0 & \mbox{if~} \lambda_{u1} = \lambda_{u2} \mbox{~and~} \lambda_{a1} =
  \lambda_{a2}.
\end{array} \right. \label{E:HamDis} 
\end{eqnarray}
That is to say, the strategy space ${\mathbb S}$ is composed of $N_c$
distinct mutually anti-correlated strategy ensemble (namely, those with same
$\lambda_{u}$); whereas the strategies of each of these ensemble are
uncorrelated with each other. (We remark that in the language of coding
theory, ${\mathbb S}$ is a linear code of $N_c^2$ elements over $GF(N_c)$
with minimum distance $L(1 - 1/N_c)$.)

We expect that the collective behavior of MG($N_c$,$N_c^2$) should follow the
predictions of the crowd-anticrowd theory as the structure of ${\mathbb S}$
matches the assumptions of the theory. In order to evaluate the
performance of players in MG($N_c$,$N_c^2$), we study the mean variance of
attendance over all alternatives (or simply the mean variance) 
\begin{equation}
 \Sigma^{2} = \frac{1}{N_c}\sum_{i=0}^{N_c} [ \langle (A_i(t))^2 \rangle -
\langle A_i(t) \rangle^2] \label{E:var},
\end{equation}
where the attendance of an alternative $A_i(t)$ is just the number of players
chosen that alternative. (We remark that the variance of
the attendance of a single alternative was studied for the MG \cite{Min1}.)  
In fact, the variance of the attendance of an
alternative represents the loss of all players in the game. The variance
$\Sigma^2$, to first order approximation, is a function of the control
parameter $\alpha$, which is the ratio of the strategy space size $|{\mathbb S}|$
to the number of strategies at play $NS$, alone \cite{Min5}.

To compare the MG($N_c$,$N_c^2$) with the crowd-anticrowd theory, we first
have to extend the calculation of the variance by the crowd-anticrowd theory
to the case of $N_c$ alternatives. According to the crowd-anticrowd theory,
the variance of the attendance originates from the independent random walk of
each mutually anti-correlated strategy ensemble. In each of these
strategy ensemble, the action of a strategy is counter-balanced by its
anti-correlated strategies. Therefore, the step size of the random walk
of a mutually anti-correlated strategy ensemble is equal to the difference
between the number of players using a single strategy from the mean number
of players using the strategies in this ensemble \cite{Min9,Min10}. 
This random walk idea can be readily extended to the case of multiple
alternatives. In fact, for the mutually anti-correlated strategy ensemble
${\mathbb S}_\lambda = \{ \lambda \vec{v}_u + \mu \vec{v}_a : \mu \in
GF(N_c) \}$,
\begin{eqnarray}
& & \mbox{step size for~} A_{\chi(\lambda,\mu)}(t)
 \mbox{~by~} {\mathbb S}_\lambda 
\nonumber \\
&=& \left| N_{\lambda,\mu} - \frac{ \sum_{\nu \in GF(N_c)} N_{\lambda,\nu}}{N_c}
 \right| \nonumber \\
&=& \frac{1}{N_c} \left| \sum_{\nu \neq \mu} (N_{\lambda,\mu} - N_{\lambda,\nu}
 ) \right| ,
\label{E:CAC_stepsize}
\end{eqnarray}
where $N_{\lambda,\mu}$ is the number of players making decision according to
the strategy $\lambda \vec{v}_u + \mu \vec{v}_a$ and $\chi(\lambda,\mu)$ is
the alternative that are chosen by the strategy $\lambda \vec{v}_u + \mu
\vec{v}_a$. Thus, the mean variance predicted by the crowd-anticrowd theory is
given by 
\begin{equation}
\Sigma^2 = \left\langle \frac{1}{N_c} \sum_{{\mathbb S}_\lambda} 
\sum_{\mu \in GF(N_c)} \left\{ \frac{1}{{N_c}^2} \left[ \sum_{\nu \neq \mu}
(N_{\lambda,\mu} - N_{\lambda,\nu}) \right]^2 \right\} \right\rangle,
\label{E:CAC_var}
\end{equation}
where $\sum_{{\mathbb S}_\lambda}$ denotes the sum of the variance over all
mutually anti-correlated strategies ensemble, and $\langle \,\rangle$ denotes
the average over time. We note that when averaged over both time and initial
choice of strategies, variance of attendance for
different alternatives must equal as there is
no preference for any alternative in the game. 

\begin{figure}[!h]
\includegraphics[scale = 0.28, bb=520 50 300 600]{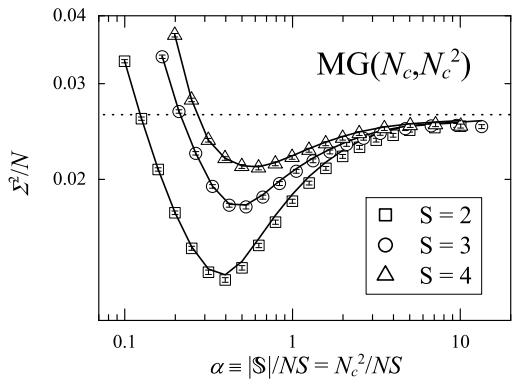}
% Here is how to import EPS art
\caption{\label{fig:f1} The mean variance $\Sigma^2$ versus the control
parameter $\alpha \equiv |{\mathbb S}|/NS = N_c^2/NS$ in MG($N_c$,$N_c^2$)
with different number of strategies $S$ where $N_c = 37$ and $M = 2$. The
solid lines are the predictions of the crowd-anticrowd theory whereas the
dashed line indicates the coin-tossed value.}
\end{figure}

Fig.~\ref{fig:f1} shows the mean variance of attendance as a function of the
control parameter $\alpha$ in the MG($N_c$,$N_c^2$) for a typical $N_c$. 
For the MG($N_c$,$N_c^2$), the mean variance of attendance, $\Sigma^2$,
exhibits similar behavior as a function of the control parameter $\alpha$ to
that in the MG no matter how many strategies $S$ players have. In particular,
whenever $N_c^2 / N S \approx 1$, the mean variance $\Sigma^2$
is smaller than the so-called coin-tossed value. (Coin-tossed value is
the mean variance resulting from players making random choices.)
Thus, global cooperation
amongst the players is observed in this parameter range. Moreover,
Fig.~\ref{fig:f1} shows that the mean variance predicted by the 
crowd-anticrowd theory agrees with our numerical finding.

Further results along this line, including the mean variance of attendance
as a function of the control parameter in MG($N_c$,$N_c^2$) with different
strategy space ${\mathbb S}$, will be reported elsewhere. These results all
agree with the crowd-anticrowd theory \cite{Min14}. Therefore, we conclude
that we have successfully build up the MG($N_c$,$N_c^2$) model whenever $N_c$
is a prime power.

Indeed, the MG($N_c$,$N_c^2$) model can be readily extended to 
MG($N_c$,$N_c^k$) with $N_c$ is equal to a prime power for $3 \le k \le M+1$.
We found that the mean variance also agrees with the MG and the crowd-anticrowd
theory in the MG($N_c$,$N_c^k$) \cite{Min14}. Thus we can always alter the
complexity of each strategy in MG with fixed $N$, $S$ and $M$ while the
cooperative behavior still persist.
\emph{As a result, we can always keep (almost) optimal cooperation amongst 
the players in almost the entire parameter space.}

However, is it possible to construct a MG with $N_c$ alternatives whose 
strategy space size is smaller than $N_c^2$ that exhibits global cooperation? 
We give the answer by constructing the MG($N_c$,$\eta N_c$) model where $\eta$
is an integer less than $N_c$. 

The basic setting of MG($N_c$,$\eta N_c$) is the same as that of
MG($N_c$,$N_c^2$) except that the strategies are drawn from a different 
strategy space. 
More precisely, strategies of MG($N_c$,$\eta N_c$) are picked from the set
${\mathbb S}_K = \{ \lambda_a \vec{v}_a + \lambda_u \vec{v}_u : \lambda_a
\in GF(N_c), \lambda_u \in K \subset GF(N_c) \}$ where $K$ contains $\eta$
elements. Moreover, $\vec{v}_a$ and $\vec{v}_u$ satisfy the two technical
conditions in Eqs.~(\ref{E:Cond_anti}) and~(\ref{E:Cond_uncorr}). Clearly,
the strategy space size of ${\mathbb S}_K$ equals $\eta N_c$. 

\begin{figure*}
\includegraphics[scale = 0.56, bb=580 40 300 580]{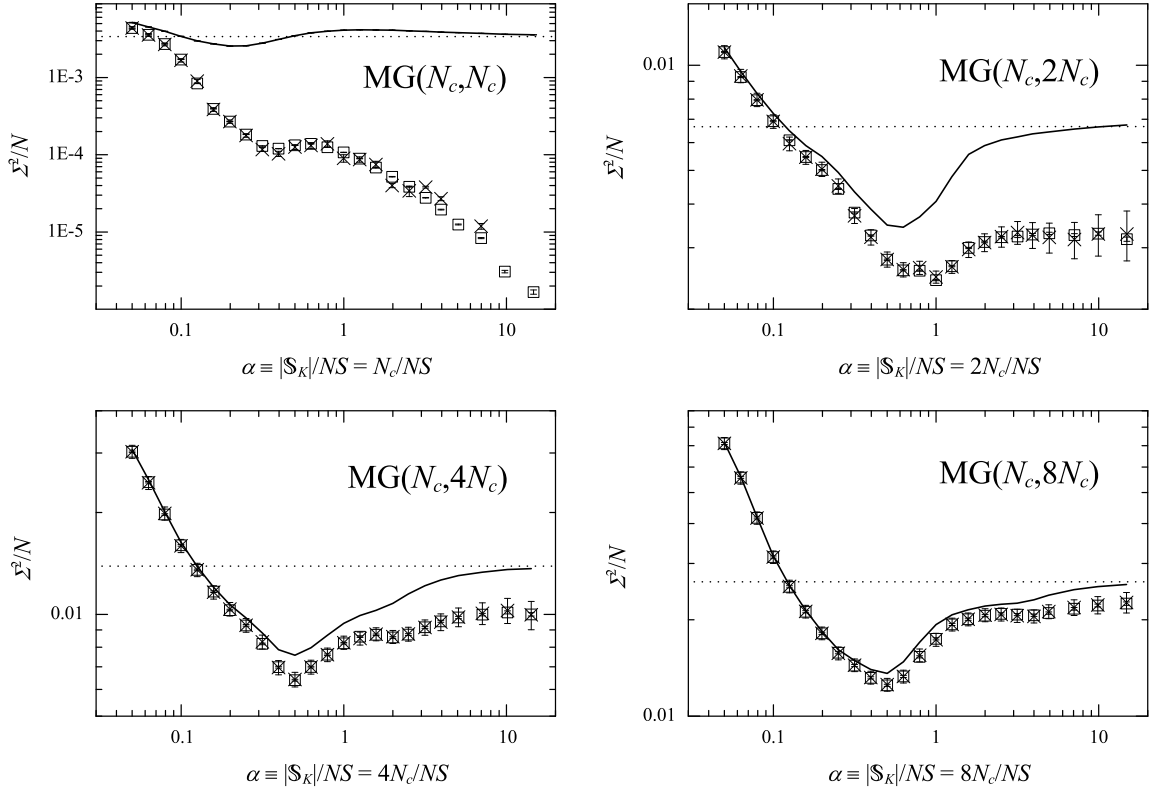}
% Here is how to import EPS art
\caption{\label{fig:f2} The mean variance $\Sigma^2$ (square) 
versus the control parameter $\alpha \equiv |{\mathbb S}_K|/NS = \eta N_c/NS$
in MG($N_c$,$\eta N_c$) with different $\eta$ where $S = 2$ and $M = 2$. 
The variance of the attendance of a choice (cross) is also shown in the figure. 
The solid line indicates the mean variance predicted by the
crowd-anticrowd theory whereas the dashed line indicates the coin-tossed value.}
\end{figure*}

As shown in Fig.~\ref{fig:f2}, the mean variance of attendance $\Sigma^2$ in
MG($N_c$,$\eta N_c$) shows similar behavior as a function of the control
parameter $\alpha$ to that in the MG only for small $\alpha$. When $\alpha$
increases, the mean variance $\Sigma^2$ in MG($N_c$,$\eta N_c$) becomes smaller
than that in MG. Nevertheless, the numerical mean variance in
MG($N_c$,$\eta N_c$) does not agree with the prediction of the crowd-anticrowd
theory except for small $\alpha$. The inconsistency is more pronounced when
$\alpha$ increases. 

To account for this discrepancy, we notice that as $\eta \rightarrow 1^+$ 
while keeping all other parameters fixed, fewer and fewer (or even none) of
the strategies in the strategy space of
MG($N_c$,$\eta N_c$) makes the same choice for the same combination of the
output of the last $M$ steps. Therefore, some of the choices can never be
chosen for MG($N_c$,$\eta N_c$) with small $\eta$ when the number of
strageties picked by the players are much smaller than the strategy space
size $\eta N_c$. In this circumstances, the attendances of most alternatives
are either one or zero. Consequently, the mean variance $\Sigma^2$ in
MG($N_c$,$\eta N_c$) with small $\eta$ is much less than $N$. In fact, the
variance of a choice may even vanished for large $\alpha$.
Such phenomenon will be more pronounced in MG($N_c$,$\eta N_c$) with small
$\eta$. Thus, the mean variance in the MG($N_c$,$\eta N_c$) for $\eta \ll N_c$
exhibits a radically different behavior from the MG.
From the above observation, we know that there is no effective crowd-anticrowd
interaction whenever $\eta \ll N_c$. And in this case, the dynamics in the
MG($N_c$,$\eta N_c$) is no longer dominated by the interactions of the
anti-correlated strategies. Consequently, the crowd-anticrowd theory does not
correctly predict the mean variance in MG($N_c$,$\eta N_c$).
Nonetheless, we still find that in MG($N_c$,$\eta N_c$), $\Sigma^2$ attains
a minimum
(and hence the average number of winning players is maximized) whenever the
control parameter $\alpha \equiv |{\mathbb S}_K|/NS = \eta N_c/NS \approx 1$
for every prime power $N_c$ and $1 < \eta < N_c$. 

Now, we are ready to answer the two questions posted in the abstract. First,
in order to obtain the best overall global wealth, players should switch to
the MG($N_c$,$\zeta N_c$) game provided that $N_c < NS \le N_c^{M+1}$. 
More specifically, for fixed $N_c$, $N$, $S$ and $M$, players simply have to
agree on an integer $\zeta \approx NS/N_c$ and the corresponding strategy space
in order to ensure the best performance of the MG. Second, whenever
$\zeta \ge N_c$, the mean variance of attendance $\Sigma^2$ agrees well with
our extension of the crowd-anticrowd theory. Thus, we conclude that in
MG($N_c$,$\zeta N_c$) with $N_c \le \zeta \le N_c^M$, the origin of global
cooperation is the self-organization of player's tendency to choose
anti-correlated strategies in making their decision. The ``cancellation'' of
the actions in these mutually anti-correlated strategy ensemble leads to a
small $\Sigma^2$.

Finally, we remark that results on the order parameter of MG($N_c$,$\zeta N_c$)
will be reported elsewhere \cite{Min14}. Readers should note that in case $N_c$
is not a prime power, the presence of zero divisors in the ring
${\mathbb Z}_{N_c}$ invalidates the conclusion in Eq.~(\ref{E:HamDis}). So, it
is instructive to find a reasonable extension of MG($N_c$,$N_c^2$) in this case. 

%\begin{acknowledgments}
Useful discussions with K.~H. Ho, P.~M. Hui and Kuen Lee is gratefully
acknowledged. This work is support by the RGC grant of the Hong Kong SAR
government under the contract number HKU~7098/00P. H.F.C. is also supported
in part by the University of Hong Kong Outstanding Young Researcher Award.
%\end{acknowledgments}

\end{document}